\documentclass[12pt]{article}
\newcommand{\GeV}{\ \textrm{GeV}}
\newcommand{\eg}{\ \textit{e.g. }}
\newcommand{\etal}{\ \textit{et al.}}
\newcommand{\np}[3]{{\it  Nucl.\ Phys.\ }{{\bf #1} {(#2)} {#3}}}
\newcommand{\pr}[3]{{\it Phys.\ Rev.\ }{{\bf #1} {(#2)} {#3}}}

\newcommand{\pl}[3]{{\it  Phys.\ Lett.\ }{{\bf #1} {(#2)} {#3}}}

\begin{document}

\title{
{\bf Q-Ball Collisions in the MSSM}}

\author{
{\sc Tuomas Multam\" aki\footnote{tuomas@maxwell.tfy.utu.fi}}\\
{\sl Department of Physics,
University of Turku} \\
{\sl FIN-20014 Turku, Finland} \\
}

\date{October 18, 2000}

\maketitle             

\abstract{Collisions of non-topological solitons, Q-balls, are studied
in the Minimal Supersymmetric Standard Model in two different cases: where
supersymmetry has been broken by a gravitationally coupled hidden sector
and by a gauge mediated mechanism at a lower energy scale. Q-ball
collisions are studied numerically on a two dimensional lattice for a
range of Q-ball charges. Total cross-sections as well as cross-sections
for fusion and charge exchange are calculated.}

\section{Introduction}
A scalar field theory with a spontaneously broken $U(1)$-symmetry
may contain stable non-topological solitons\cite{leepang,coleman},
Q-balls. A Q-ball is a coherent state of a complex scalar field
that carries a global $U(1)$ charge. In the sector of fixed charge
the Q-ball solution is the ground state so that its stability and 
existence are due to the conservation of the $U(1)$ charge.
In realistic theories Q-balls are generally allowed in supersymmetric 
generalizations of the standard model with flat directions in their
scalar potentials. Q-balls have been shown to be present in the 
MSSM\cite{kusenko2,enqvist1} where leptonic and baryonic balls may exist and 
they may be formed in the early universe by a mechanism
that is closely related to the Affleck-Dine 
baryogenesis\cite{enqvist1,kusenko3,kawasaki1,kawasaki2}.

Stable Q-balls can contribute to the dark matter content of 
the universe. These can be balls with charges of the order of
$10^{20}$ but also very small Q-balls can be considered
as dark matter\cite{kusenko3,kusennko2}. On the other hand
decaying Q-balls can protect baryons from the erasure of 
baryon number due to sphaleron transitions by decaying after the
electroweak phase transition\cite{enqvist1}.
Q-ball decay may also result in the production of dark matter
in the form of the lightest supersymmetric particle (LSPs). This process 
may explain the baryon to dark matter ratio of the universe\cite{enqvistdm}.

\section{Q-ball solutions}
Consider a field theory with a U(1) symmetric scalar potential 
$U(\phi)$ that has a global minimum at $\phi=0$ and the complex scalar field
$\phi$ carries a unit charge with respect to the $U(1)$-symmetry.
The charge and energy of a field configuration $\phi$ are given
by
\begin{equation}
Q={1\over i}\int (\phi^*\partial_t\phi-\phi\partial_t\phi^*)d^3x
\end{equation}
\begin{equation}
E=\int d^3x[|\dot{\phi}|^2+|\nabla\phi|^2+U(\phi)].
\end{equation}

The Q-ball solution is the minimum energy configuration at a fixed
charge. If it is energetically favourable to store
charge in a Q-ball compared to free particles, the Q-ball will be stable.
Hence for a stable Q-ball, condition $E<mQ$, where $m$ is the mass of
the $\phi$-scalar, must hold.

The Q-ball solution can be shown to be of the form 
$\phi(x,t)=e^{i\omega t}\phi(r)$,
where $\phi(r)$ is now time independent, spherically symmetric and real.
$\omega$ is the so called Q-ball frequency, $\omega\in [-m,m]$. Q-balls
can be characterized by the value of $\omega$: the larger
$\omega$ the larger the charge carried by the Q-ball.

\section{Q-ball profiles}
Q-balls and their cosmological significance have been studied
in the MSSM mainly in two different types of potentials that correspond 
to SUSY broken by a gravity- or gauge-mediated mechanism. 

The potentials in these two cases are respectively
\begin{eqnarray}
U_{Gr}(\phi) & = & m_1^2\phi^2(1-K \log({\phi^2\over M^2}))+\lambda_1
\phi^{10}\\
U_{Ga}(\phi) & = & m_2^4(1+\log({\phi^2\over m^2}))+{\lambda_2^2\over m_{Pl}^2}
\phi^6.
\end{eqnarray}
The parameter values we have chosen are:
$m_1=10^2\GeV,\ K=0.1,\ \lambda_1=m_{Pl}^{-6},\ m_2=10^4\GeV,\ \lambda_2=0.5$
The large mass scale, $M$, is chosen such that the minimum is degenerate.

Q-ball profiles are of different type in these two cases. In the 
gravity-mediated case the profiles are well approximated by a Gaussian
ansatz. The radius of a Q-ball is only weakly dependent on charge
and Q-balls are typically thick-walled. In the gauge-mediated case the
profile of a ball is more dependent on charge and as charge increases the
Q-ball profiles become thin-walled. 

\begin{figure}[ht]
\leavevmode
\centering
\vspace*{50mm}
\includegraphics{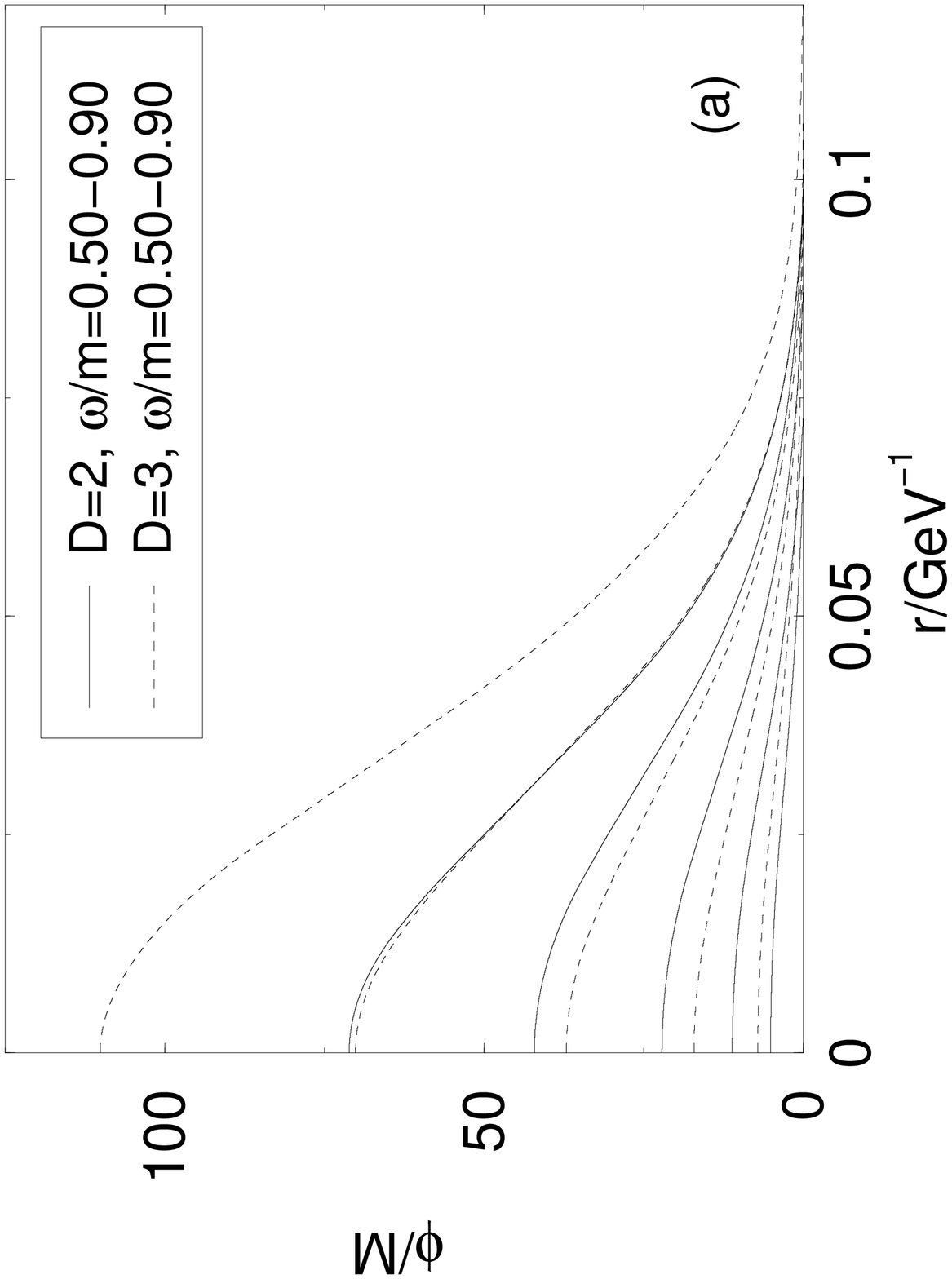}
\includegraphics{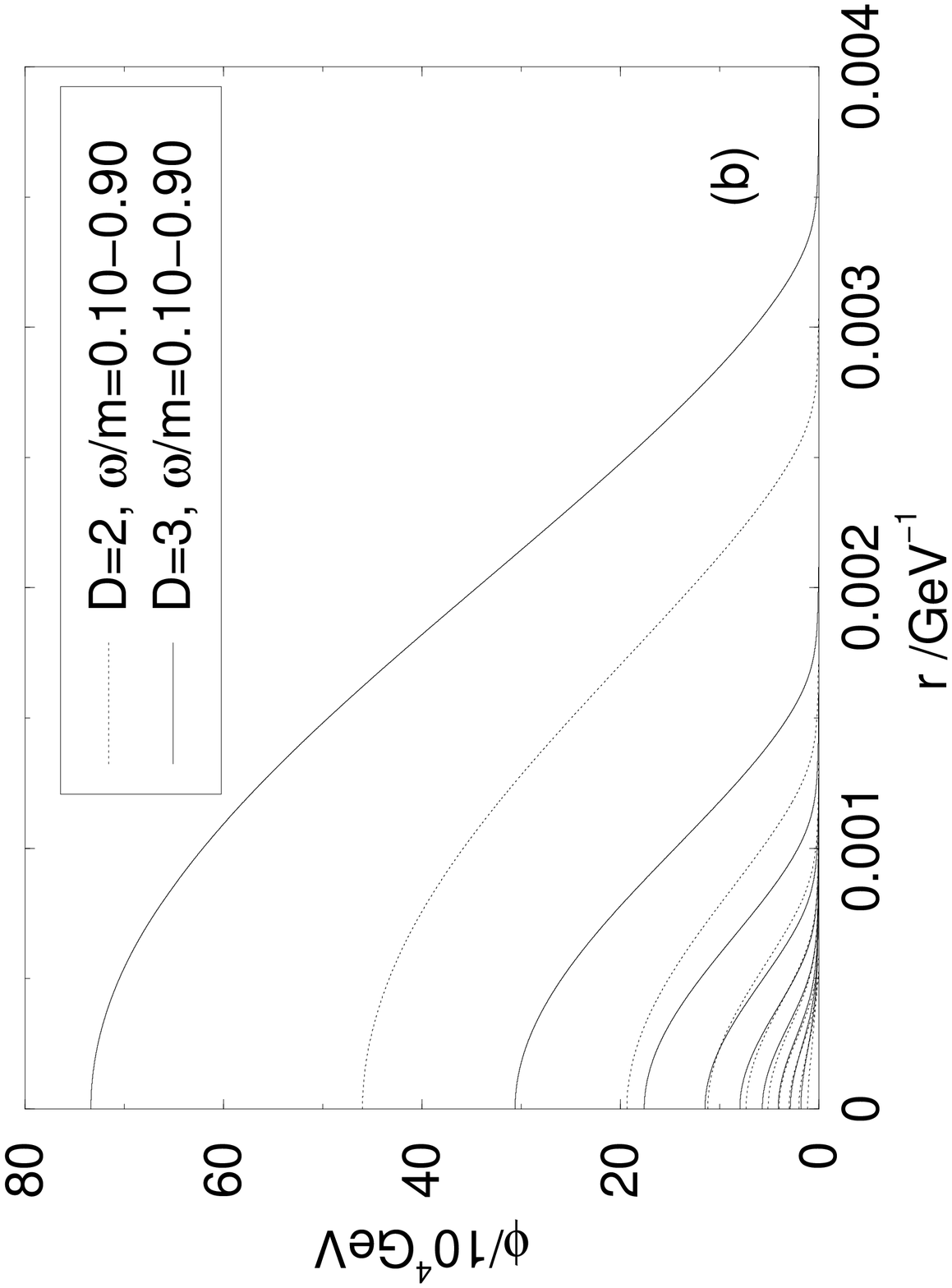}
\caption{Q-ball profiles in the (a) gravity- and (b) gauge-mediated case in two
and three dimensions}\label{profs}
\end{figure}

Q-ball profiles in the two cases have been plotted in figure \ref{profs}
in two and three dimensions for different values of $\omega$. 

\section{Collisions}
Collisions of Q-balls may play an important role in their cosmological
history. After the Q-balls have been formed, collisions can alter the charge
distribution significantly which in turn may have an effect
on the importance of Q-balls to cosmology, \eg if Q-balls 
fragment due to collisions they may evaporate\cite{cohen} 
and cannot be responsible for the dark matter content of the universe. 
Q-ball collisions have been previously considered 
analytically\cite{kusenko3}
and numerically\cite{axenides,belova,makhanov,battye} 
in various potentials. To address the cosmological issues, however, 
one needs to calculate cross sections in realistic potentials.

We have studied Q-ball collisions in the gravity- and gauge-mediated
cases\cite{multamak2,multamak3} on a $(2+1)$-dimensional lattice for
a range of Q-ball sizes (charges). As from figure \ref{profs} can be
seen, the Q-ball profiles in two and three dimensions are similar
and one can expect that the results obtained in these two dimensional 
simulations well approximate the results from the more realistic
three dimensional calculations.

Collisions have been studied for a range of charges, relative phase
differences, $\Delta\phi$, and different velocities. In each case the 
colliding balls have equal charges.
Collisions in both potentials have similar features: 
when the relative phase difference is small the balls fuse. 
In a fusion process some charge is typically lost as small lumps
and radiation. As $\Delta\phi$
increases the balls start to scatter while charge is transferred from
one ball to the other. As $\Delta\phi$ grows further the amount of 
transferred charge decreases until the Q-balls scatter elastically.

\begin{figure}[ht]
\leavevmode
\centering
\vspace*{55mm}
\includegraphics{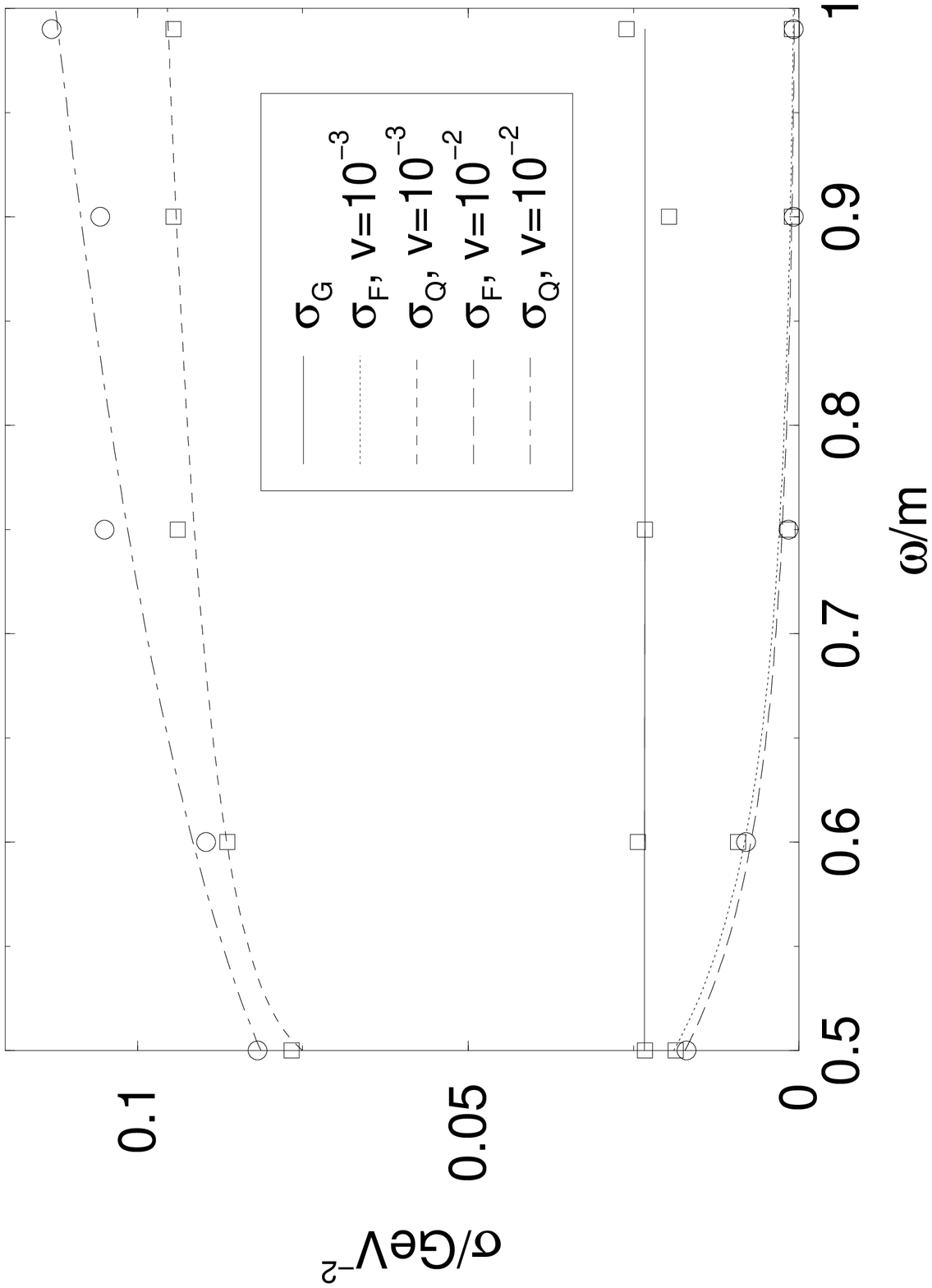}
\caption{Geometric, fusion and charge-exchange cross-sections for
$v=10^{-3}$ and $v=10^{-2}$ in the gravity-mediated scenario}
\label{gravcross}
\vspace*{100mm}
\includegraphics{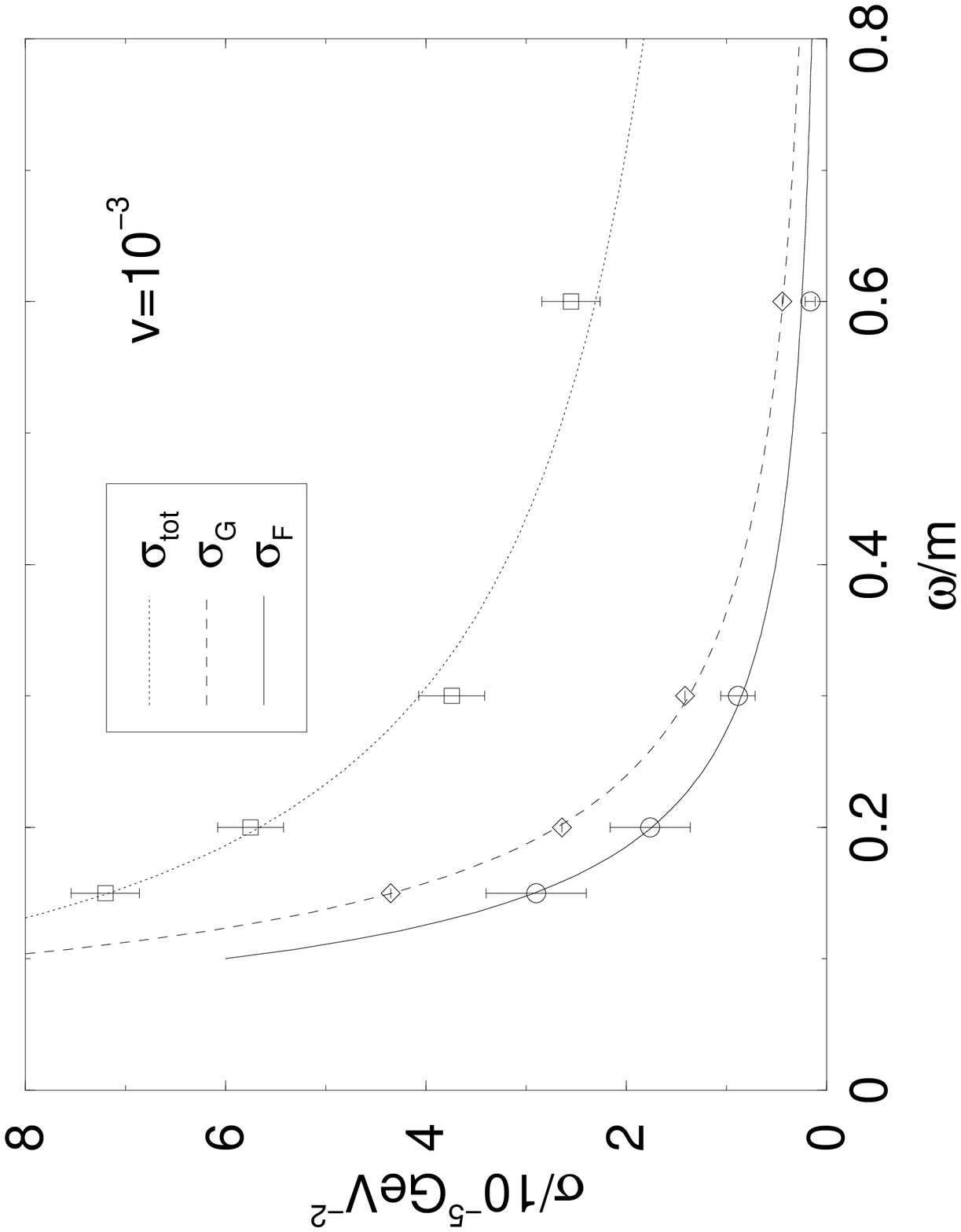}
\includegraphics{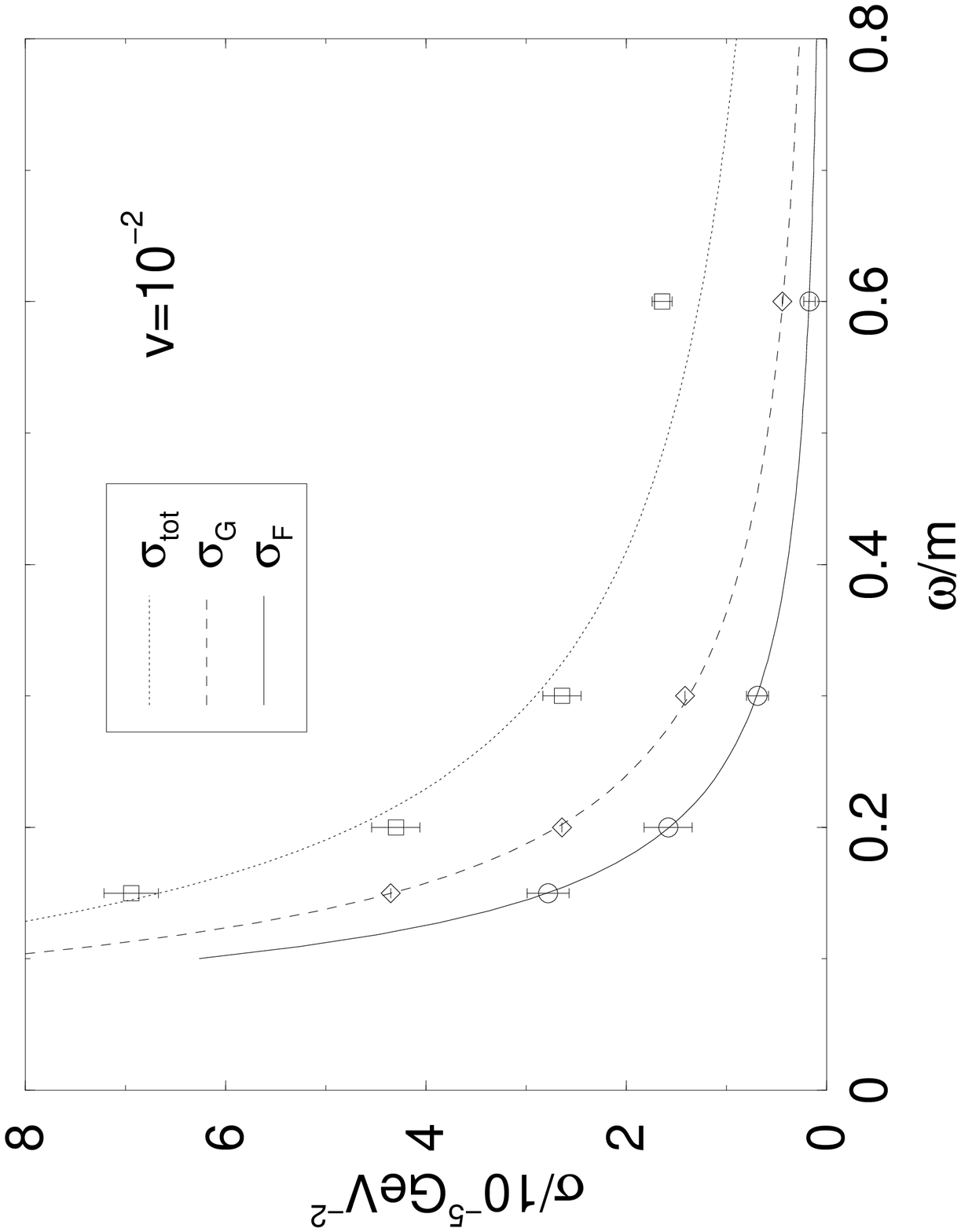}
\caption{Total, geometric and fusion cross-sections for
$v=10^{-3}$ and $v=10^{-2}$ in the gauge-mediated scenario}
\label{gaugecross}
\end{figure}

By varying the impact parameter the cross-sections for each process
and the total cross-section have been calculated. The total, 
$\sigma_{tot}$, geometric, $\sigma_G$, fusion, $\sigma_F$,
and charge-exchange, $\sigma_Q$, cross-sections averaged 
over the relative phase have been plotted in figures 
\ref{gravcross} and \ref{gaugecross} for different $\omega$:s
in the gravity- and gauge-mediated scenarios.

\section{Conclusions}
Q-ball collisions have been studied in the MSSM with supersymmetry
broken by two different mechanisms. Even though the Q-balls in the
two cases have differing characteristics, the qualitative features
of a collision process are alike: Q-balls either fuse, exchange charge
or scatter elastically. In both cases the relative phase
difference between the colliding Q-balls determines the type of the 
collision.

\section*{Acknowledgments}
We thank the Finnish Center of Scientific Computing for computation
resources. This work has been supported by the Finnish Graduate School
in Nuclear and Particle Physics.



\begin{thebibliography}{X}
\bibitem{leepang} T. Lee and Y. Pang, \pr{221}{1992}{251}.
\bibitem{coleman} S. Coleman, \np{B262}{1985}{263}.
\bibitem{kusenko2} A. Kusenko, \pl{B404}{1997}{285}.
\bibitem{enqvist1} K. Enqvist and J. McDonald, \pl{B425}{1998}{309}.
\bibitem{kusenko3} A. Kusenko and M. Shaposhnikov, \pl{B418}{1998}{46}.
\bibitem{kawasaki1} S. Kasuya and M. Kawasaki, \pr{D61}{2000}{041301}.
\bibitem{kawasaki2} S. Kasuya and M. Kawasaki, \pr{D62}{2000}{023512}.
\bibitem{kusennko2} A. Kusenko, \pl{B405}{1997}{108}.
\bibitem{enqvistdm} K. Enqvist and J. McDonald, \np{B538}{1999}{321}.
\bibitem{cohen} A. Cohen \etal, \np{B272}{1986}{301}.
\bibitem{axenides} M. Axenides \etal, \pr{D61}{2000}{085006}.
\bibitem{belova} T. Belova and A. Kudryavtsev, \textit{Zh. Eksp. Teor. Fiz.}
{\bf 95} (1989) 13.
\bibitem{makhanov} V. Makhankov \textit{et al.}, \pl{A70}{1979}{171}.
\bibitem{battye} R. A. Battye and P. M. Sutcliffe, hep-th/0003252.
\bibitem{multamak2} T. Multam\"aki and I. Vilja, \pl{B482}{2000}{161}.
\bibitem{multamak3} T. Multam\"aki and I. Vilja, \pl{B484}{2000}{283}.
\end{thebibliography}
\end{document}